\documentclass[prl,twocolumn,superscriptaddress,showpacs]{revtex4}
\usepackage{amssymb}
\usepackage[tbtags]{amsmath}
\usepackage{graphicx}

\newcommand{\ket}[1]{| #1 \rangle}

\newcommand{\eq}[1]{(\ref{#1})}

\DeclareMathOperator{\re}{Re}

\begin{document}

\title{Tunable coupling of superconducting qubits}
\author{Alexandre Blais}
\email{alexandre.blais@yale.edu}
\altaffiliation[present address: ]{Department of Physics, Yale University, P.O. Box 208120, New Haven, CT 06520-8120}
\affiliation{D\'epartement de Physique and Centre de Recherche sur les
Propri\'et\'es \'Electroniques de Mat\'eriaux Avanc\'es,\\
Universit\'e de Sherbrooke, Sherbrooke, Qu\'ebec, J1K 2R1, Canada}
\author{Alexander \surname{Maassen van den Brink}}
\email{alec@dwavesys.com}
\affiliation{D-Wave Systems Inc., 320-1985 West Broadway, Vancouver, B.C., V6J 4Y3, Canada}
\author{Alexandre M. Zagoskin}
\email{zagoskin@dwavesys.com}
\affiliation{D-Wave Systems Inc., 320-1985 West Broadway, Vancouver, B.C., V6J 4Y3, Canada}
\affiliation{Physics and Astronomy Dept., The University of British Columbia, 
6224 Agricultural Rd., Vancouver, B.C., V6T 1Z1, Canada}
\date{\today}

\begin{abstract}
We study an $LC$-circuit implemented using a current-biased Josephson 
junction (CBJJ) as a tunable coupler for superconducting qubits.
By modulating the bias current, the junction can be tuned in and out of
resonance and entangled with the qubits coupled to it. One can thus implement two-qubit operations by mediating entanglement. We consider the examples of CBJJ and charge--phase qubits.  A simple recoupling scheme leads to a generalization to arbitrary qubit designs.
\end{abstract}
\maketitle

Significant successes in manipulating the quantum state of
superconducting qubits~\cite{vion:2002,martinis:2002,han:2002,pashkin:2002} once more
make them prime candidates for a solid-state quantum computer~\cite{makhlin:2001}.  Since the experiments yield single-qubit coherence times close to the accepted limits~\cite{devoret:90}, one can focus on other steps towards realizing the potential of quantum information processing~\cite{nielsen-chuang} in these systems. The critical next step is controlled coupling of, at least, two qubits.

Several coupling mechanisms are possible, e.g., capacitive coupling~\cite{martinis_pers} for charge, charge--phase~\cite{vion:2002}, and current-biased Josephson-junction (CBJJ) qubits~\cite{martinis:2002,han:2002}. Importantly, it is simple to implement and recently enabled entangling two charge qubits~\cite{pashkin:2002}. Also, this type of coupling can be turned on and off by tuning the qubits' level spacings in and out of resonance [if the interaction Hamiltonian is off-diagonal in the computational basis, e.g.\ $\sigma_x{\otimes}\sigma_x$]. A clear disadvantage is that tuning the qubits themselves may cause extra decoherence.  Moreover, not all qubits are thus tunable, or have off-diagonal interactions. To avoid this problem, the coupling can be controlled using refocusing pulses---similar to liquid-state NMR, where the $J$-coupling must be refocused~\cite{slichter}. In this case, universal quantum computing is still possible, but imperfect refocusing introduces errors and the threshold for fault tolerance is not yet known.

We propose to capacitively couple superconducting qubits to a CBJJ, implementing an $LC$-circuit and acting as a tunable bus. This parallels cavity QED~\cite{raimond:2001} (the CBJJ and qubits playing the roles of the cavity and atoms respectively) and ion traps~\cite{CiracZoller95}. $LC$-circuits can also be coupled to flux qubits (inductively)~\cite{anatoly:2002} and other superconducting devices~\cite{buisson:2000,marquandt:2001,al-saidi:2001}. In another scheme to entangle qubits through an $LC$-circuit~\cite{makhlin:2001}, the latter's virtual states mediate an effective qubit--qubit interaction.

The CBJJ's kinetic inductance depends on the bias and modifies the
circuit's overall inductance~\cite{Ilichev2002}. It thus
acts as a tunable \emph{an}harmonic $LC$-circuit, which guarantees a
non-uniform level spacing, reducing leakage to higher
states. However, for anharmonic oscillators, transitions from $|n\rangle$ to $|n{\pm}2\rangle$ 
etc.\ can cause leakage; this is minimized for suitable system parameters (see below).

\begin{figure}[tbp]
\includegraphics[width=3in]{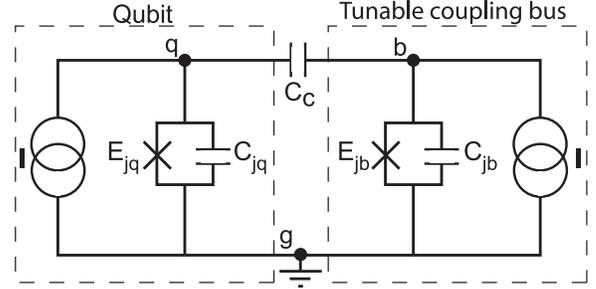}
\caption{A pair of capacitively coupled CBJJ qubits.}
\label{fig_qubit_coupler}
\end{figure}

For illustration, we first consider a pair of CBJJs, coupled by a capacitance $C_\mathrm{c}$ (Fig.~\ref{fig_qubit_coupler}; cf.\ Ref.~\cite{Johnson}). One plays the role of the qubit and the other of the tunable bus. Controlled coupling of charge--phase qubits follows. Coupling of a charge qubit to a CBJJ was studied in Ref.~\cite{hekking:2002}.

A Josephson junction biased by a dc current has the well-known
washboard potential~\cite{clarke:1988}. Close to the critical bias~$I_\mathrm{c}$,
there are few levels in each washboard well. We consider a large junction with a bias such that there are only three such levels~\cite{martinis:2002}. Then, the two lowest levels in one of the wells form the qubit's computational subspace $\{|0\rangle ,|1\rangle \}$. State $|n\rangle$ has decay rate $\Gamma_n$ to the continuum. The $| 0 \rangle\leftrightarrow| 1 \rangle$ transition frequency is~$\Omega$.

The circuit of Fig.~\ref{fig_qubit_coupler} has the Hamiltonian
\begin{equation}\begin{split}
\mathcal{H} &=
\frac{p_\mathrm{q}^{2}}{2\tilde{C}_\mathrm{jq}} - E_\mathrm{jq}\cos \phi _\mathrm{q}
- \frac{\Phi _{0}}{2\pi }I_\mathrm{q}\phi _\mathrm{q}\\
&\quad
+ \frac{p_\mathrm{b}^{2}}{2\tilde{C}_\mathrm{b}}
- E_\mathrm{jb}\cos\phi_\mathrm{b} - \frac{\Phi_{0}}{2\pi}I_\mathrm{b}\phi_\mathrm{b}
+\frac{p_\mathrm{q}p_\mathrm{b}}{\tilde{C}_\mathrm{c}}\;;
\end{split}\label{H_circuit}
\end{equation}
$p_{i}$ is the charge at node $i$ and $\phi _{i}$ the phase difference
across junction $i$; `q' (`b') denotes qubit (bus). The effective capacitances are
$\tilde{C}_\mathrm{jq} =C_\mathrm{jq}+(C_\mathrm{jb}^{-1}{+}C_\mathrm{c}^{-1})^{-1}$, $\tilde{C}_\mathrm{jb} =C_\mathrm{jb}+(C_\mathrm{jq}^{-1}{+}C_\mathrm{c}^{-1})^{-1}$, and $\tilde{C}_\mathrm{c} =C_\mathrm{jq}C_\mathrm{jb}(C_\mathrm{jq}^{-1}{+}C_\mathrm{jb}^{-1}{+}C_\mathrm{c}^{-1})$. 
Below we take both junctions identical: 
$C_\mathrm{jq}=C_\mathrm{jb}\equiv C_\mathrm{j}$ and $E_\mathrm{jq}=E_\mathrm{jb}\equiv E_\mathrm{j}$.

For near-critical bias, the washboard potentials are well approximated by cubic
ones~\cite{leggett:87} and the junctions can be treated as anharmonic oscillators.
Using this analogy, the charge at node $i$ is
$p_i = i\frac{2\pi}{\Phi_0}\sqrt{m\hbar\omega_\mathrm{p}/2}\,(a_i^\dag-a_i)$,
with the `mass' $m=\tilde{C}_\mathrm{j}(\Phi_0/2\pi)^2$ and the plasma frequency $\omega_{\mathrm{p}i}=\sqrt{2\pi I_\mathrm{c}/\tilde{C}_\mathrm{j}\Phi_0}[1-(I_{\mathrm{bias},i}/I_\mathrm{c})^2]^{1/4}$; $a_i^{(\dag)}$ is an annihilation (creation) operator \cite{leggett:87,clarke:1988}.

Expressing (\ref{H_circuit}) in the basis $\{| 0_\mathrm{q} \rangle,| 1_\mathrm{q} \rangle,|
2_\mathrm{q} \rangle\}\otimes\{| 0_\mathrm{b} \rangle,| 1_\mathrm{b} \rangle,| 2_\mathrm{b} \rangle\}$, we find the coupled eigenstates. First, focus on the Hamiltonian $\mathcal{H}_2$ in $\mathcal{L}$, the span of $\{| 0_\mathrm{q}1_\mathrm{b} \rangle,| 1_\mathrm{q}0_\mathrm{b} \rangle\}$: to first order in the anharmonicity,
\begin{equation}
  \mathcal{H}_2 = \left(
  \begin{array}{cc}
  E_\mathrm{q0}+E_\mathrm{b1} & \gamma/2 \\
  \gamma/2 & E_\mathrm{q1}+E_\mathrm{b0}
  \end{array}\right),  \label{H1orderBloc}
\end{equation}
where the coupling coefficient is
$\gamma\equiv\hbar \sqrt{\omega_\mathrm{pq}\omega_\mathrm{pb}}\*\,\tilde{C}_\mathrm{j}/ \tilde{C}_\mathrm{c}$ and $E_{ik}$ is the energy of level $k$. Without coupling, $| 0_\mathrm{q}1_\mathrm{b}\rangle$ and $| 1_\mathrm{q}0_\mathrm{b} \rangle$ are degenerate for bias currents such that\linebreak $E_\mathrm{q1}-E_\mathrm{q0} = E_\mathrm{b1}-E_\mathrm{b0}$. A non-zero $\gamma$ lifts the degeneracy, and the new (maximally entangled) eigenstates are
\begin{equation}
  \ket{\psi_\pm}\equiv(| 0_\mathrm{q}1_\mathrm{b} \rangle\pm
   | 1_\mathrm{q}0_\mathrm{b} \rangle)/\sqrt{2}\;.\label{eigenstates}
\end{equation}

In resonance, $\mathcal{H}_2$ thus acts as $e^{-i\sigma_{\!x}\!\gamma\tau\!/2\hbar}$ in $\mathcal{L}$ and as phase factors outside. Hence, for a system prepared in $| 1_\mathrm{q}0_\mathrm{b}\rangle$, the probability to find the qubit in $| 1_\mathrm{q} \rangle$ oscillates with period $T_\mathrm{Rabi}=h/\gamma$. For a single CBJJ such oscillations only occur under current bias at frequency~$\Omega$ \cite{martinis:2002,han:2002}. Oscillations for the coupled qubit and bus without applied resonant perturbation on them individually then demonstrate their entanglement.

Anharmonicity is crucial here as it keeps other relevant level pairs out of resonance, suppressing leakage out of~$\mathcal{L}$. However, $p_\mathrm{q}p_\mathrm{b}/\tilde{C}_\mathrm{c}$ in (\ref{H_circuit}) causes non-resonant leakage, in particular to $| 2_\mathrm{q(b)} \rangle$. These states are closer to the top of the potential barrier, so $\Gamma_{2,\mathrm{q(b)}}$ are large. Hence, poisoning of $\ket{\psi_\pm}$ with $| 2_\mathrm{q(b)} \rangle$ shortens the coherence time.

We evaluate the extent of this leakage numerically. The Hamiltonian of each anharmonic oscillator is expressed in terms of about $20$ harmonic-oscillator eigenstates and diagonalized. For each junction, we use $C_\mathrm{j}=6$pF and $I_\mathrm{c}=21\mu$A~\cite{martinis:2002}. With $I_{\mathrm{bias}}=20.8\mu $A, each well contains three levels. We take $C_\mathrm{c}=25$fF, minimizing leakage while keeping a reasonable $T_\mathrm{Rabi}\approx40$ns (see below)~\cite{martinis:2002}.

We find that $|2_\mathrm{q(b)}\rangle$ poisons $\ket{\psi_\pm}$ with a small probability ($P_2\sim10^{-6}$) only, as expected. The coupled system's other states also have weak poisoning by $|2_\mathrm{q(b)}\rangle$; worst is the eigenstate close to $|1_\mathrm{q}1_\mathrm{b}\rangle$, with $P_2\sim10^{-4}$. Roughly, the lifetime of a state with poisoning $P_2$ is $(P_2\Gamma_{2})^{-1}$. Since $\Gamma_{2}/\Gamma_{1}\sim 10^{3}$~\cite{martinis:2002}, then $P_2\lesssim10^{-4}$ should hardly change the lifetime of the qubit or bus.

When the bus is not tuned to the qubit's frequency~$\Omega_\mathrm{q}$, the two are decoupled. For this, we keep the qubit's bias constant, and for the bus decrease it to 20.43$\mu$A; each well then contains about 11 levels. The eigenstates are now computed as $0.007 | 0_\mathrm{q}1_\mathrm{b} \rangle +0.999 | 1_\mathrm{q}0_\mathrm{b} \rangle$ and $0.999| 0_\mathrm{q}1_\mathrm{b} \rangle + 0.007 | 1_\mathrm{q}0_\mathrm{b} \rangle$, where poisoning by higher states with probabilities \mbox{$\lesssim10^{-6}$} has been omitted.

\begin{figure}[tbp]
\includegraphics[width=3in]{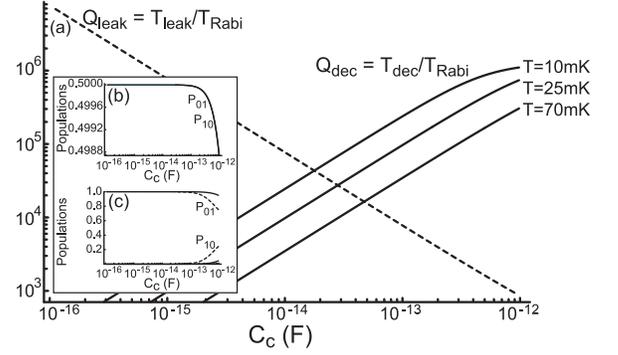}
\caption{Quality of coupled identical CBJJ qubits. (a)~Decoherence time $T_\mathrm{dec}$ (full lines) and leakage time $T_\mathrm{leak}$ (dashed line) over the oscillation period $T_\mathrm{Rabi}$, vs the coupling capacitance~$C_\mathrm{c}$. We take $T_\mathrm{dec}^{-1}=T_1^{-1}+T_2^{-1}$ [see Eqs.\ \eq{eq_T1} and~\eq{eq_T2} below] and $T_\mathrm{leak}^{-1}=\Gamma_2(1{-}P_{01}{-}P_{10})$; $P_{ij}$ is the population of $\ket{i_\mathrm{q}j_\mathrm{b}}$. The resonant cases ($I_{\mathrm{bias}}=20.8\mu $A) at $T=10$, $25$, and 70mK are shown. At 25mK~\cite{martinis:2002}, $C_\mathrm{c}\sim10$fF maximizes the effective quality factor $Q$. (b)~Populations $P_{01}$ and $P_{10}$ of $|\psi_+\rangle$ vs $C_\mathrm{c}$ in resonance. At large $C_\mathrm{c}$, poisoning by other states reduces $P_{01}$ and $P_{10}$. (c)~Same as (b) but off resonance: $I_{\mathrm{bias,q}}=20.8\mu $A, $I_{\mathrm{bias,b}}=20.43\mu $A (full lines) and $I_{\mathrm{bias,b}}=20.74\mu $A (dashed lines).}
\label{fig_numerical}
\end{figure}

One can choose a $C_\mathrm{c}$ optimizing the effective quality of the coupled qubits, as shown in Fig.~\ref{fig_numerical}. To avoid further leakage and gate errors, the qubit--bus coupling should be turned on faster than $T_\mathrm{Rabi}$ but adiabatically with respect to the bus interlevel spacing, corresponding to ${\sim}1$ns~\cite{martinis:2002}; this leaves a suitable window of turn-on time.

\begin{figure}[tbp]
\vspace*{0.1cm} \includegraphics[width=3.25in]{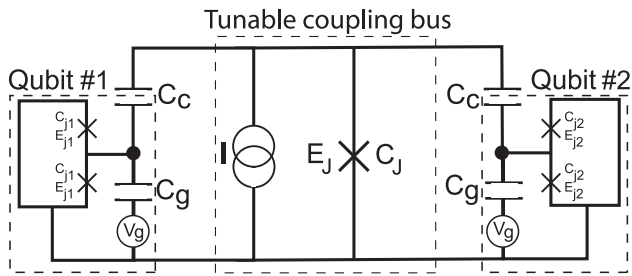}
\caption{A pair of charge--phase qubits capacitively coupled to a CBJJ.}
\label{fig_S-JJ-S}
\end{figure}

Let us turn to a pair of charge--phase qubits coupled through a CBJJ (Fig.~\ref{fig_S-JJ-S}).  For the qubits, only two levels are considered. To be able to couple the bus to only one qubit at a time, we assume $\Omega_1\ne\Omega_2$ for their level spacings. Similarly to the above, tuning the bus in resonance with $\Omega_i$ causes coherent oscillations between it and qubit~$i$, while the other qubit is hardly affected.

As before, the interaction Hamiltonian couples the bus charge to the qubit-island charge and, in the logical basis for the qubit~\cite{vion:2002}, takes the form $\sigma_{ix}p_\mathrm{b}/\tilde{C}_\mathrm{c}$. The qubit--bus coupling coefficient is
$\gamma' \equiv \sqrt{2}\beta(2e)\*(2\pi/\Phi_0)^2\*\sqrt{m\hbar\omega_\mathrm{pb}}/ \tilde{C}_\mathrm{c}$, where $\tilde{C}_\mathrm{c}$ now depends on the total capacitances of both qubits $C_{\Sigma,i}=C_{\mathrm{g}i}+2C_{\mathrm{j}i}$. The number $\beta$ depends on the ratio of the qubit island's charging and Josephson energies; we take $\beta = 1.16$, corresponding to the parameters of Ref.~\cite{vion:2002}.

For two-qubit operations, the qubits should interact sequentially with the bus which, at the end of the operation, should be disentangled from them. This can be done as follows. Assume the qubits are in an arbitrary state and the bus prepared in its ground state. The bus is first tuned to $\Omega_1$ for a time $t_1$ such that $\gamma' t_1/2\hbar = \pi/2$. It is then tuned to $\Omega_2$ for a time $t_2$ with $\gamma' t_2/2\hbar = \pi/4$ and, finally, tuned again to $\Omega_1$ for another~$t_1$. Afterwards, the bus is disentangled from the qubits. Omitting some phase factors, the net effect is to implement a square-root of swap on the qubits. Together with single-qubit operations, this gate is universal for quantum  computation~\cite{loss:98}. The phase factors have to be accounted for. Since all energies involved are known from numerics and depend on experimentally accessible parameters, this should not be a problem. Moreover, since these qubits always have $\Omega_i\ne0$, they accumulate phase shifts. Refocusing on the idle qubit is therefore assumed.

As above, leakage can occur to higher bus states. Taking, e.g., $I_\mathrm{c}=147.9\mu$A, $C_\mathrm{j}=5.8$pF~\cite{han:2002}, and $I_{\mathrm{bias}}\gtrsim 0.99 I_\mathrm{c}$, then $\Omega_\mathrm{b}$ is in the range of the qubit energy splitting in Ref.~\cite{vion:2002}. For the qubits, we take $C_\Sigma=5.5$fF, and $\Omega_1$ and $\Omega_2$ equal to $\Omega_\mathrm{b}$ at $I_{\mathrm{bias}} = 146.5\mu$A and $146.75\mu$A respectively. Further, $C_\mathrm{c} =0.1$fF. These values reduce mixing while keeping the coherent oscillations as fast as possible, $h/\gamma'\sim100$ns, of the order of the single-qubit Rabi period under microwave excitation in Ref.~\cite{vion:2002}.

Since charge--phase qubits are at least as anharmonic as CBJJs, the leakage per qubit will be no larger than for the circuit in Fig.~\ref{fig_qubit_coupler}. With two qubits, the total Hilbert space is however larger, leaving more room for leakage. A higher density of states also means that operations must be slower to avoid spurious transitions, hence the longer Rabi period. As above, poisoning with $| 2_\mathrm{b} \rangle$ has probability $P_2\sim10^{-4}$ and smaller $C_\mathrm{c}$ helps to avoid mixing, but at the price of a longer Rabi period. [A small $C_\mathrm{c}$ also ensures that the qubit islands' charging energies are virtually unchanged.] Moreover, the qubit--bus eigenstates are not maximally entangled as in~(\ref{eigenstates}), their amplitudes having a small difference ${\sim}10^{-3}$. This will have to be accounted for when realizing logic operations. Finally, biasing the bus at $I_\mathrm{bias} = 146.6\mu$A decouples it from both qubits, again with $P_2\sim10^{-4}$.

Besides leakage, other imperfections must be dealt with, in particular relaxation and dephasing due to fluctuations of the control parameters---$I_\mathrm{bias}$, gate voltage, etc. The effect of these noises on a single qubit or CBJJ has been given in Refs.\ \cite{makhlin:2001,martinis:2002b}. Here we study them for the coupled system near the degeneracy point. Consider, e.g., the CBJJs in Fig.~\ref{fig_qubit_coupler}. Major sources of decoherence are the qubit $\delta I_\mathrm{q}(t)$ and bus $\delta I_\mathrm{b}(t)$ bias noises. These correspond to two separate environments, leading to independent fluctuations of $\omega_\mathrm{pq}$ and $\omega_\mathrm{pb}$. Near degeneracy, focus again on the subspace $\mathcal{L}$. First rewrite (\ref{H1orderBloc}) as 
$\mathcal{H}_2 = \frac{1}{2}\epsilon(\sigma_z\cos\theta+\sigma_x\sin\theta)$, where 
$\cos\theta = (\Omega_\mathrm{q}{+}\Omega_\mathrm{b})/\epsilon$, $\sin\theta = \gamma/\epsilon$, and $\epsilon=\sqrt{(\Omega_\mathrm{q}{+}\Omega_\mathrm{b})^2+\gamma^2}$. Then, expand $\mathcal{H}_2$ to $\mathcal{O}(\delta I)$, obtaining the system--bath Hamiltonian
$\mathcal{H}_{\mathrm{SB}} = \tau_z (X_\mathrm{q}{+}X_\mathrm{b}) + \tau_x (X'_\mathrm{q}{+}X'_\mathrm{b})$. The $\tau_n$ are Pauli matrices in the eigenbasis of $\mathcal{H}_2$ and $X^{(\prime)}_i$ are bath operators.

For an environment represented by real impedances $Z(\omega)=R_{I}$ in parallel to each junction, the spectral density of the bath operators is
$J(\omega)=J_{x}(\omega)+J_{z}(\omega)$, where
\begin{equation}
\begin{split}
\frac{R_{I}J_{z}(\omega )}{\omega }= & \biggl[ \frac{\omega _\mathrm{pq}I_\mathrm{bias,q}
}{4(I_\mathrm{cq}^2{-}I_\mathrm{bias,q}^2)}\left( \frac{\gamma\sin
\theta }{2\omega _\mathrm{pq}}-\hbar \cos \theta \right) \biggr] ^{2}  \\
& +\biggl[\frac{\omega _\mathrm{pb}I_\mathrm{bias,b}}{%
4(I_\mathrm{cb}^{2}{-}I_\mathrm{bias,b}^{2})}\left( \frac{\gamma\sin \theta
}{2\omega _\mathrm{pb}}+\hbar \cos \theta \right) \biggr] ^{2}
\end{split}
\end{equation}
and similarly for $J_x(\omega)$ but with $\theta\mapsto\theta+\pi/2$. The effects of the two
independent baths add up in $J(\omega )$.

The relaxation and dephasing rates now become~\cite{grifoni:99}
\begin{align}
  T_{1}^{-1} &= J_{x}(\epsilon/\hbar)\coth(\epsilon / 2k_\mathrm{B}T)/2\hbar\;,
\label{eq_T1}\\
  T_{2}^{-1} &= T_1^{-1}\!/2+2\pi \alpha k_\mathrm{B}T/\hbar\;,
\label{eq_T2}
\end{align}
where $\alpha = J_{z}(\omega)/2\pi\hbar\omega$ for $\omega\rightarrow 0$. For the CBJJ, long coherence times require an environment with sufficiently high impedance, engineered to be $\re Z(\omega )\approx560\mathrm{k}\Omega$ in Ref.~\cite{martinis:2002} (instead of the standard ${\sim}100\Omega$ at microwave frequencies). With this $R_{I}$, $T=25$mK, and the above parameters for two CBJJs, $T_{1,2}$ are ${\sim}1$ms at exact resonance. Out of resonance, $T_{1}$ grows while $T_{2}$ stays of the same order. This lower bound on $T_{1,2}$ is $>10^3\times$ the decoherence time of the decoupled system \cite{vion:2002,martinis:2002,han:2002}, so $\delta I_\mathrm{q,b}$ should hardly affect the coherence.

For a charge--phase qubit at the working point, voltage fluctuations
$\delta V$ across the coupling-bus impedance $Z_\mathrm{b}(\omega)$
affect the qubit island through $\mathcal{H}_\mathrm{SB}\sim e\beta\delta
V\*(C_\mathrm{c}/C_\Sigma)\tau'_x$.
Here, $\tau'_x$ is a Pauli matrix in the single-qubit eigenbasis. As is
clear from the general \eq{eq_T1} and~\eq{eq_T2}, there is no dephasing
since $\mathcal{H}_\mathrm{SB}$ does not couple to
$\tau'_z$~\cite{vion:2002}.  Moreover, in the case of a purely resistive
$Z_\mathrm{b}(\omega)$, and for now considering only the junctions'
capacitance~\cite{makhlin:2001}, the spectrum of $\delta V$ is centered
at $\omega=0$.  With, e.g., a large
$\re Z_\mathrm{b}(\omega)\approx560\mathrm{k}\Omega$ as above, this
spectrum is very narrow and its weight at $\Omega_{1,2}$ small. The
relaxation due to $Z_\mathrm{b}$ is therefore weak~\cite{blais:2002}.

Finally, let us generalize to other qubit designs. The above holds if the interaction Hamiltonian is completely off-diagonal in the computational basis. If it is of the form $\sigma_{z\mathrm{q}}{\otimes}q_\mathrm{b}$, recoupling can be used~\cite{lidar:2002}. Start from the relations $H\sigma_xH=\sigma_z$ and $H\sigma_zH=\sigma_x$ for the Hadamard gate $H$~\cite{nielsen-chuang}. Since $H_\mathrm{q}(B_{x\mathrm{q}}\sigma_{x\mathrm{q}} + \Omega_\mathrm{b}\sigma_{z\mathrm{b}} + \frac{1}{2}\gamma\sigma_{z\mathrm{q}}{\otimes}p_\mathrm{b})H_\mathrm{q} = B_{x\mathrm{q}}\sigma_{z\mathrm{q}} + \Omega_\mathrm{b}\sigma_{z\mathrm{b}} + \frac{1}{2}\gamma\sigma_{x\mathrm{q}}{\otimes}p_\mathrm{b}$, applying $H$ on the qubit before and after the qubit--bus interaction (realized here by taking the single-qubit Hamiltonian as $B_{x\mathrm{q}}\sigma_{x\mathrm{q}}$ with $B_{x\mathrm{q}}=\Omega_\mathrm{b}$) the coupled system will behave as if it had an off-diagonal interaction. All our results then apply.

In conclusion, we have considered a CBJJ acting as a tunable $LC$-circuit to mediate
entanglement between superconducting qubits and perform logic
operations. The method allows coupling qubits not just with
different parameters, but of different kinds: the two qubits in
Fig.~\ref{fig_S-JJ-S} need not be the same. It allows to switch
the coupling on and off without tuning individual qubits.
Estimates show that no significant additional relaxation or
dephasing is introduced into the system. Leakage to higher states
can be minimized by choosing the junction
parameters and coupling capacitances. The issue of leakage has been
addressed previously~\cite{fazio:99},
and this should be adapted to apply here. Since the qubits and
$LC$-circuit can be optimized independently, this gives reason for optimism. Our approach
thus has the potential to lead to a universal 
coupling scheme for solid-state qubit registers.

As this article was being finalized, a paper investigating
capacitive coupling between charge qubits through an
$LC$-circuit appeared on the LANL preprint
server. In Ref.~\cite{Plastina2002}, coupling between the qubits is realized using
both on- and off-resonant pulses with the $LC$-circuit.

\begin{acknowledgments}
AB thanks J.M. Martinis for discussions and for
giving a copy of his work prior to publication. We
thank M.H.S. Amin, J. Gallop, A. Golubov, J.P. Hilton, E.~Il'ichev, S. Lacelle, S. Marchand, A.N. Omelyanchouk, A.Yu.\ Smirnov, A.M.
Tremblay, and A.~Tzalenchuk for discussions and
remarks on the manuscript. AB was supported by NSERC, FCAR,
D-Wave Systems Inc., and IMSI.
\end{acknowledgments}

\end{document}